\documentclass[12pt]{article}
\newcommand{\Comment}[1]{{}}
\usepackage{amsfonts,amsthm,amsmath,amssymb,slashed}
\usepackage[textwidth = 430 pt, textheight = 630 pt]{geometry}

\Comment{\usepackage{color}
\definecolor{MyDarkBlue}{rgb}{0.15,0.15,0.45}
\usepackage[linktocpage=true]{hyperref}
\hypersetup{
colorlinks=true,
citecolor=MyDarkBlue,\bibliography{heisenberg}
linkcolor=MyDarkBlue,
urlcolor=MyDarkBlue,
pdfauthor={Horatiu Nastase},
pdftitle={DBI scalar field theory for QGP hydrodynamics},
pdfsubject={hep-th}
}

\usepackage[numbers,sort&compress]{natbib}
\usepackage{hypernat}}
\usepackage{graphicx}

\newcommand\ignore[1]{}
\def\one{{\,\hbox{1\kern-.8mm l}}}

\def\a{\alpha}\def\b{\beta}

\def\d{\partial}

\newcommand{\Cset}{{\,\,{{{^{_{\pmb{\mid}}}}\kern-.45em{\mathrm C}}}}}

\newcommand{\be}{\begin{equation}}
\newcommand{\bea}{\begin{eqnarray}}

\newcommand{\ee}{\end{equation}}
\newcommand{\eea}{\end{eqnarray}}

\begin{document}

\renewcommand{\thefootnote}{\fnsymbol{footnote}}

\makeatletter
\@addtoreset{equation}{section}
\makeatother
\renewcommand{\theequation}{\thesection.\arabic{equation}}

\rightline{}
\rightline{}
   \vspace{1.8truecm}


\vspace{10pt}


\begin{center}
{\LARGE \bf{\sc  DBI scalar field theory for QGP hydrodynamics}}
\end{center}
 \vspace{1truecm}
\thispagestyle{empty} \centerline{
{\large \bf {\sc Horatiu Nastase}}\footnote{E-mail address: \Comment{\href{mailto:nastase@ift.unesp.br}}{\tt
    nastase@ift.unesp.br}}
}

\vspace{1cm}

\vspace{.8cm}
\centerline{{\it 
Instituto de F\'{i}sica Te\'{o}rica, UNESP-Universidade Estadual Paulista}} \centerline{{\it
R. Dr. Bento T. Ferraz 271, Bl. II, Sao Paulo 01140-070, SP, Brazil}}

\vspace{1.0truecm}

\thispagestyle{empty}

\centerline{\sc Abstract}

\vspace{.4truecm}

\begin{center}
\begin{minipage}[c]{380pt}
{\noindent A way to describe the hydrodynamics of the quark-gluon plasma using a DBI action is proposed, 
based on the model found by Heisenberg for high energy scattering of nucleons. 
The expanding plasma is described as a shockwave in a DBI model for a real scalar standing in for the pion, and I show that one obtains a fluid 
description in terms of a relativistic fluid that near the shock is approximately ideal ($\eta\simeq 0$) and conformal. 
One can introduce an extra term inside the square root of the DBI action that generates a shear viscosity term in the 
energy-momentum tensor near the shock, as well as a bulk viscosity, and regulates the behaviour of the energy density at the shock, 
making it finite. The resulting fluid satisfies the relativistic Navier-Stokes equation with $u^\mu, \rho,P,\eta$ defined in terms of $\phi$ and 
its derivatives.
One finds a relation between the parameters of the theory and the QGP thermodynamics, $\a/\b^2=\eta/(sT)$, and 
by fixing $\a$ and $\b$ from usual (low multiplicity) particle scattering, one finds $T\propto m_\pi$. 
}
\end{minipage}
\end{center}

\vspace{.5cm}

\setcounter{page}{0}
\setcounter{tocdepth}{2}

\newpage

\tableofcontents
\renewcommand{\thefootnote}{\arabic{footnote}}
\setcounter{footnote}{0}

\linespread{1.1}
\parskip 4pt


\section{Introduction}

One of the most interesting phenomena in high energy physics being studied today is the creation of a strongly coupled Quark-Gluon Plasma (sQGP)
obtained in high energy heavy ion collisions \cite{Adams:2005dq,Adcox:2004mh,Back:2004je,Arsene:2004fa,Gyulassy:2004zy}. 
It is known that the expanding plasma is well described by an (effective) hydrodynamics description, 
with a shear viscosity $\eta$ over entropy density $s$ close \cite{Romatschke:2007mq,Heinz:2013th}
to the result for generic gravity duals with black holes, $\eta/s=1/(4\pi)$, \cite{Policastro:2001yc,Buchel:2003tz}. For a while it was thought that this value
was a lower bound for any theory \cite{Kovtun:2004de}, but then it was realized that there can be violations \cite{Kats:2007mq}. 
Moreover, the plasma then decays into tens of thousands of particles, with a constant "freeze-out" temperature of about $175 MeV$
\cite{BraunMunzinger:2003zd,BraunMunzinger:2004eg,BraunMunzinger:2001ip}. 

The hydrodynamics is an effective description, in an expansion in derivatives on a velocity field, for a strongly coupled quantum field theory. But at high enough 
energies and for high enough multiplicities of the emitted particles, one can have a classical field theory description as well, in terms of
some effective action. This is the basis of the model that Heisenberg introduced in 1952 \cite{Heisenberg1952}
in order to describe the asymptotic high energy scattering of 
nucleons. Even though this was before QCD, and thus also before the discovery of the Froissart unitarity bound \cite{Froissart:1961ux,Lukaszuk:1967zz}
for high energy scattering, 
\be
\sigma_{\rm tot}(s)\leq C\log^2(s/s_0);\;\;\; C\leq \frac{\pi}{m_\pi^2}\;,
\ee
whose saturation is notoriously difficult to obtain in effective models for nonperturbative QCD (the saturation should happen in a highly nonperturbative
regime, of very high multiplicity for the emission of the lowest energy QCD particles, the pions), the Heisenberg model easily predicts the saturation 
of the Froissart bound at high enough energies,
\be
\sigma_{\rm tot, Heisenberg}(s)\simeq \frac{\pi}{m_\pi^2}\log^2\frac{s}{s_0}.
\ee
The effective action that Heisenberg considered is of the DBI type, for a scalar pion field, and the saturation of the Froissart bound is obtained
in an analysis of the particle emission from a classical shockwave solution of the DBI action, combining the classical field theory analysis with 
some quantum mechanics, namely the idea that the classical field is made of quanta of energy. 

It was shown in \cite{Kang:2005bj} that the picture of Heisenberg is obtained directly from an analysis of the gravity dual scattering, using 
the set-up for high energy scattering of Polchinski and Strassler \cite{Polchinski:2001tt} for AdS/CFT \cite{Maldacena:1997re}. The work 
was based on the gravity dual calculation of the scattering in the gravity dual in \cite{Kang:2004jd}, making precise the earlier result
\cite{Giddings:2002cd}. Recently, the Heisenberg model was revisited in \cite{Nastase:2015ixa}, to analyze its possible applications, generalizations and 
relation to gravity duals. 

In this paper I consider the DBI action of Heisenberg as a good effective description of the sQGP near the shock in the shockwave solution, 
and I show that we can obtain a hydrodynamics description that matches the features observed in the sQGP. The DBI action leads to 
an approximately ideal relativistic fluid with $P\simeq \rho/3$ and $\eta\simeq 0$. One can introduce an extra term inside the square root in
the DBI action, that leads to a finite nonzero $\eta/s$ near the shock, as well as a bulk viscosity $\zeta=2\eta/3$. 
This term then regulates the behaviour of the energy density near the 
shock, which would diverge in its absence, thus one can argue that it is needed for physical reasons. 
A first attempt to derive a hydrodynamics from the DBI action was presented in section 5 of \cite{Nastase:2005pb}.

The resulting action has two free parameters, that are a priori free, one is the DBI parameter (in string theory related to $\a'$), and the other is the 
coefficient of the $\eta$ term. It is unclear how one could fix them simply from considerations of sQGP hydrodynamics, but one can instead consider
the DBI action as also an action relevant for low multiplicity (usual) particle scattering, i.e. as a quantum effective action, of a chiral perturbation 
theory type. Then we can fix the parameters from experiments, in terms of $f_\pi$ and $m_\pi$. We find a relation between the parameters of
the theory and its thermodynamics, and together with fixing the parameters in terms of $f_\pi$ and $m_\pi$, we will find a prediction for the 
QGP thermodynamics.

The paper is organized as follows. In section 2 I review the Heisenberg model for nucleon-nucleon scattering, and in section 3 I review 
relativistic hydrodynamics. In section 4 I present the hydrodynamics of the DBI scalar, first considering the subtleties of the free massless
scalar in section 4.1, then describing the ideal DBI hydrodynamics in section 4.2. In section 4.3 I introduce the viscous term, in section 4.3 
I show that it regulates the behaviour near the shock, and in section 5 I conclude.

\section{Heisenberg model}

Heisenberg considered that at high enough energy, where the Froissart saturation regime dominates, the nonperturbative physics is well described by a 
DBI scalar action with a mass term inside the square root, 
\be
S_{\rm DBI}=-\b^{-2}\int d^4x \left[\sqrt{1+\b^2[(\d_\mu\phi)^2+m^2\phi^2]}-1\right].\label{dbi}
\ee

In collisions of hadrons at high enough energy, one can consider that the hadrons and the pion field around them are Lorentz contracted, appearing as a 
shockwave in the limit. Thus effectively, one can consider collisons of pion field shockwaves described by the DBI scalar. 

To describe the shockwave, one considers a simple 1+1 dimensional model, with a spatial coordinate $x$ and temporal coordinate $t$. Relativistic invariance 
(and the fact that the shockwave in this ultrarelativistic regime moves at the speed of light) dictates the dependence $\phi=\phi(s)$ only, with 
\be
s=t^2-x^2
\ee
and no dependence on the transverse coordinates $y,z$.
Note that just relativistic invariance would imply dependence on both $x^+=t+x$ and $x^-=t-x$, but adding the fact that $x^-=0$ or $x^+=0$ must be 
the location of the shock in every Lorentz frame uniquely selects the dependence on $x^+x^-=s$. 

The equations of motion of (\ref{dbi}) are
\bea
&&\d_\mu\frac{\d_\mu \phi}{\sqrt{1+\b^2[(\d_\mu\phi)^2+m^2\phi^2]}}-\frac{m^2\phi}{\sqrt{1+\b^2[(\d_\mu\phi)^2+m^2\phi^2]}}=0\Rightarrow\cr
&&-\Box \Phi+m^2\phi+\b^2\frac{(\d_\mu\d_\nu\phi)(\d_\mu\phi)\d_\nu \phi}{1+\b^2[(\d_\mu\phi)^2+m^2\phi^2]}=0.
\eea

On the ansatz $\phi=\phi(s)$, we have
\be
(\d_\mu\phi)^2=-4d\left(\frac{d\phi}{ds}\right)^2;\;\;\;
\Box \phi=-4\frac{d}{ds}\left(s\frac{d\phi}{ds}\right)\;,
\ee
so the Lagrangean on the solution becomes
\be
{\cal L}=-\b^{-2}\left[\sqrt{1+\b^2\left[-4s\left(\frac{d\phi}{ds}\right)^2+m^2\phi^2\right]}-1\right]\;,
\ee
and the equation of motion becomes, after cancelling some terms and rewriting,
\be
4\frac{d}{ds}\left(s\frac{d\phi}{ds}\right)+m^2\phi=8s\b^2\left(\frac{d\phi}{ds}\right)^2\frac{\frac{d\phi}{ds}+m^2\phi}{1+\b^2m^2\phi^2}.
\ee

The perturbative solution at small $s$ is, independent on whether $m=0$ or not, 
\be
\phi\simeq \frac{\sqrt{s}}{\b}+...
\ee
Then $\phi'(s)\rightarrow \infty$ at $s\rightarrow 0$, and
\be
\b^2(\d_\mu\phi)^2\rightarrow -1\;,
\ee
which means that the square root in the DBI action goes to zero. 

Moreover, the usual energy-momentum tensor defined by coupling to gravity (the Belinfante tensor) of the DBI scalar is
\be
T^B_{\mu\nu}
=-\frac{2}{\sqrt{-g}}\frac{\delta S}{\delta g^{\mu\nu}}=\frac{\d_\mu\phi\d_\nu\phi-g_{\mu\nu}(\d_\rho\phi)^2-g_{\mu\nu}\b^{-2}(1+\b^2m^2\phi^2)}{\sqrt{1
+\b^2[(\d_\mu\phi)^2+m^2\phi^2]}}+\frac{g_{\mu\nu}}{\b^2}\;,\label{belinDBI}
\ee
so we see that the energy density $T_{00}$ goes to infinity at the shock, since the square root vanishes there. 

As far as Heisenberg's model for the saturation of the Froissart bound goes, the value of $\b$ cannot be fixed by experiments. 
It is also hard to see how we could fix the parameter from the hydrodynamics description in section 4. 
But if we also consider 
the model as a chiral perturbation theory model for the pion, we can. Indeed then, expanding the square root to second order, we get 
\be
S_{\rm DBI}=\int d^4x \left[-\frac{1}{2}(\d_\mu\phi)^2-\frac{1}{2}m^2\phi^2+\frac{b^2}{8}[m^4\phi^4+2m^2\phi^2(\d_\mu\phi)^2+((\d_\mu\phi)^2)^2]\right]+...
\ee
But in general, the coefficient of the $\phi^2(\d_\mu\phi)^2$ term is $f_\pi^{-2}/6$, so we identify 
\be
\b=\frac{\sqrt{2}}{\sqrt{3}m_\pi f_\pi}\simeq\frac{1}{(126MeV)^2}\;,
\ee
since $m_\pi\simeq 140 MeV$ and $f_\pi \simeq 93MeV$. Of course, the parameter $m$ in the DBI action is identified with $m_\pi\simeq 140MeV$.

\section{Relativistic hydrodynamics}

In relativistic hydrodynamics, for an isotropic fluid with energy density $\rho$ and pressure $P$, one expands the energy 
momentum tensor in terms of derivatives
acting on the 4-velocity $u^\mu$ as 
\bea
T_{\mu\nu}&=&\rho u_\mu u_\nu +P(g_{\mu\nu}+u_\mu u_\nu)+\pi_{\mu\nu}\cr
\pi_{\mu\nu}&=&-2\eta\left[\frac{\nabla_\mu u_\nu+\nabla_\nu u_\mu}{2}+\frac{a_\mu u_\nu+a_\nu u_\mu}{2}
-\frac{1}{3}(\nabla^\rho u_\rho)(g_{\mu\nu}+u_\mu u_\nu)\right]\cr
&&-\zeta(\nabla^\mu u_\mu)
(g_{\mu\nu}+u_\mu u_\nu)+...\;,
\eea
where $a_\mu=u^\rho \nabla_\rho u_\mu$, 
the expansion was written only up to first order in derivatives, was written in the Landau frame $u^\mu \pi_{\mu\nu}=0$, and $\eta$ is the 
shear viscosity, and $\zeta$ is the bulk viscosity. 

The equation of motion of the fluid, the equivalent of the Navier-Stokes equation at the relativistic level, is given by substituting the 
expansion into the conservation equation 
\be
\d^\mu T_{\mu\nu}=0\;,
\ee
giving at zeroth order
\be
u_\mu u_\nu \d^\mu(\rho+P)+(\rho+P)(\d^\mu u_\mu) u_\nu +(\rho+P)u_\mu \d^\mu u_\nu +\d_\nu P=0.
\ee
It is easy to see that in the nonrelativistic limit $u^\mu\simeq (1,\vec{v})$, $P\ll \rho$, separating the equation into $\nu=0$ and $\nu=i$ components and
after some algebra, we obtain the usual nonrelativistic equations, Euler's equation and the continuity equation, respectively,
\bea
\rho\frac{\d \vec{v}}{\d t}+\rho(\vec{v}\cdot\vec{\nabla})\vec{v}+\vec{\nabla}P&=&0\cr
\d_t\rho+\vec{\nabla}\cdot(\rho\vec{v})&=&0.
\eea

At next order, we obtain a relativistic generalization of the Navier-Stokes equation,
\bea
&&u_\mu u_\nu \d^\mu(\rho+P)+(\rho+P)(\d^\mu u_\mu) u_\nu +(\rho+P)u_\mu \d^\mu u_\nu +\d_\nu P\cr
&=&\eta \d^\mu(\d_\mu u_\nu +\d_\nu u_\mu)+\left(\zeta-\frac{2\eta}{3}\right)\d^\mu\left[(\d^\rho u_\rho)(g_{\mu\nu}+u_\mu u_\nu)\right]\;,\label{relNavSto}
\eea
which in the nonrelativistic limit gives the Navier-Stokes equation instead of Euler's equation,
\be
\rho\frac{\d \vec{v}}{\d t}+\rho(\vec{v}\cdot\vec{\nabla})\vec{v}+\vec{\nabla}P=\eta\Delta \vec{v}+\left(\zeta-\frac{2\eta}{3}\right)\vec{\nabla}(\vec{\nabla}
\cdot\vec{v}).
\ee

We are especially interested in conformal fluids, i.e. fluids for conformal invariant systems. Conformal invariance means we should be able to find a 
traceless energy-momentum tensor, and in particular for the above energy-momentum tensor, $T_\mu^\mu=0$  leads to 
\be
\rho=3P;\;\; \zeta=0.
\ee

But there is a subtlety. Only in $d=2$ does conformal invariance imply a traceless Belinfante tensor, i.e. the standard energy-momentum tensor obtained
by coupling to gravity. Indeed, only there conformal invariance implies Weyl invariance, which leads to a traceless Belinfante tensor. 

In $d=4$, that is no longer true. In fact, for the simplest conformally invariant object we can think of, a free massless scalar field, the Belinfante tensor 
is not traceless,
\bea
S&=&\int d^4x \left[-\frac{1}{2}(\d_\mu \phi)^2\right]\Rightarrow\cr 
T_{\mu\nu}^B&\equiv& -\frac{2}{\sqrt{-g}}\frac{\delta S}{\delta g^{\mu\nu}}=
\d_\mu \phi\d_\nu\phi-\frac{1}{2}g_{\mu\nu}[(\d_\rho\phi)^2]\Rightarrow\cr
{T^{B\mu}}_\mu&=&\frac{2-d}{2}(\d_\rho\phi)^2.
\eea

But that is easily fixed. One can use the Noether ambiguity, that one can alwasy add the divergence of a current to the energy-momentum tensor
without changing the conservation law $\d^\mu T_{\mu\nu}=0$, 
\be
T_{\mu\nu}\rightarrow T_{\mu\nu}+\d_\lambda J^{\mu\nu\lambda}\;,
\ee
where $J^{\mu\nu\lambda}=-J^{\lambda\nu\mu}$, to define an improved energy-momentum tensor 
\be
T^I_{\mu\nu}=T^B_{\mu\nu}-\frac{1}{6}(\d_\mu \d_\nu -g_{\mu\nu}\d^2)\phi^2\;,\label{improvedfree}
\ee
that is now traceless on-shell,
\be
{T^{I\mu}}_\mu={T^{B\mu}}_\mu+(\d_\mu\phi)^2+\phi\d^2\phi\;,
\ee
so when using the equation of motion $\d^2\phi=0$, we obtain zero. 

So, if we want to derive an (approximately) conformal fluid out of an (approximately) 
conformally invariant field theory, we must use an improved energy-momentum tensor, not the Belinfante tensor.

\section{DBI scalar hydrodynamics}

We want to understand the approximately ideal fluid that is created in heavy ion collisions: the state of matter dubbed strongly-coupled 
Quark-Gluon Plasma (sQGP) is known to be described by an approximately ideal hydrodynamics, with $\eta/s\sim 1/(4\pi)$ (which was thought for a 
time to be the absolute minimum value the ratio can have, for any system), i.e. with a very small shear viscosity, as well as being approximately 
conformal, i.e. $P=\rho/3$ at leading order, perhaps also with vanishing bulk viscosity at the next order, $\zeta\simeq 0$, though 
recently it has been argued that perhaps the bulk viscosity is of the same order as the shear viscosity \cite{Ryu:2015vwa}. We will 
see that we are led to naturally have $\zeta\sim \eta$, more specifically $\zeta=2\eta/3$.

This state is created in high energy collisions in the high multiplicity regime (creation of many pions) responsible for the saturation of the Froissart 
bound for $\sigma_{\rm tot}(\tilde s)$. But in this regime the Heisenberg model works very well, so we are led to believe that the approximately
ideal fluid hydrodynamics should be also described by it. 

Of course, we have presented only a very simple one-dimensional model, with a shock propagating on a null line, $s=0$, so we will not be able to 
extract too much information from it. 

The first question to ask is, if the scalar $\phi(s)$ describes a fluid, how to define the 4-velocity $u^\mu$? The first hint comes from the fact that 
for a irrotational nonrelativistic flow, one can find a potential $\Phi$ such that $\vec{v}=\vec{\nabla}\Phi$. The second hint comes from the fact that, if 
we expand in momenta the scalar field,
\be
\phi(x)=\int d^4k e^{ik\cdot x}\phi(k)\Rightarrow \d^\mu \phi\sim ik^\mu \phi\;,
\ee
so for $\phi(k)$ peaked on a single value, we would have the 4-velocity of the field be $u^\mu=k^\mu/m\propto \d^\mu\phi$. We therefore define 
the 4-velocity in general as
\be
u^\mu=\frac{\d^\mu\phi}{\sqrt{-(\d_\mu\phi)^2}}\;,
\ee
which has the desired property that $u^\mu u_\mu=-1$, provided that $(\d_\mu\phi)^2<0$. This is however correct on the shock solution $\phi(s)$, 
as we have already seen. 

Note that this definition is consistent with, and thus is a generalization of, the usual definition for the density and pressure of a canonically 
normalized scalar field with a potential, $S=\int d^4x\sqrt{-g}[-1/2(\d_\mu\phi)^2-V(\phi)]$, in the case when the field is only time-dependent, 
$\phi=\phi(t)$. Our definition implies then $u^\mu=(1,0,0,0)$, and thus $T_{\mu\nu}={\rm diag}(\rho,p,p,p)$, i.e. a homogeneous isotropic ideal fluid,
and on the other hand from the Belinfante tensor one obtains
\be
\rho_\phi=\frac{1}{2}\dot\phi^2+V(\phi);\;\;\;
p_\phi=\frac{1}{2}\dot\phi^2-V(\phi)\;,
\ee
and the other components of $T_{\mu\nu}$ are zero, consistent with our definition. We can therefore think of our definition as the natural relativistic 
generalization of this standard case.

\subsection{Free massless scalar}

But we want to recover a conformal fluid for the (conformal) case of the free massless scalar field, so we must use an improved energy-momentum 
tensor. If we use the improved energy-momentum tensor (\ref{improvedfree}), which gives on-shell the traceless result
\be
T^I_{\mu\nu}=\frac{2}{3}\d_\mu\phi\d_\nu\phi-\frac{1}{6}g_{\mu\nu}(\d_\rho\phi)^2-\frac{\phi}{3}\left(\d_\mu\d_\nu\phi
-\frac{1}{4}g_{\mu\nu}\d^2\phi\right)\;,
\ee
then the first two terms can be identified with an ideal fluid term, and the second with a shear viscosity term (the first order term in the derivative 
expansion). This gives the ideal fluid parameters
\bea
P+\rho&=&-\frac{2}{3}(\d_\mu\phi)^2\cr
P&=&-\frac{1}{6}(\d_\mu\phi)^2\;,
\eea
which leads to 
\bea
\rho&=&-\frac{1}{2}(\d_\mu\phi)^2\cr
P&=&-\frac{1}{6}(\d_\mu\phi)^2\;,
\eea
satisfying the conformal fluid relation $P=\rho/3$. Moreover, if we consider a solution $\phi=\phi(s)$, which for a massless scalar implies
\be
-4\frac{d}{ds}\left(s\frac{d\phi}{ds}\right)=0\Rightarrow \phi(s)=C\log s\;,
\ee
then $(\d_\mu\phi)^2=-4s \phi'^2<0$, so $\rho>0, P>0$. In turn, that means that
\be
(\d_\mu\phi)^2=-\frac{4C^2}{s}\rightarrow -\infty\;\;\;{\rm as}\;\; s\rightarrow 0\;,
\ee
so the energy density blows up at the shock, $\rho\rightarrow \infty$. 

We can also extract information about the next order in the expansion. We first note that 
\be
\d_\mu u_\nu=\frac{1}{\sqrt{-(\d\phi)^2}}\left[\d_\mu\d_\nu\phi- \d_\nu\phi \frac{(\d_\mu\d^\rho\phi)\d_\rho\phi}{(\d_\lambda\phi)^2}\right]\;,
\ee
so considering the term with two derivatives on $\phi$ in the energy-momentum tensor as the viscous term $\pi_{\mu\nu}^{(1)}$, 
we find the shear viscosity as
\be
2\eta=\frac{\phi}{3}\sqrt{-(\d_\lambda\phi)^2}\;,
\ee
and $\zeta=0$ as wanted for a conformal fluid. Then $\eta$ is finite as $s\rightarrow 0$, however
\be
\frac{2\eta}{\rho+P}=\frac{\phi}{2\sqrt{-(\d_\lambda\phi)^2}}\rightarrow 0\;\;{\rm as}\;\; s\rightarrow 0.\label{etas}
\ee
Note that we described the whole fluid as a function of $\phi$, with $u^\mu\propto \d^\mu\phi$ and $\rho\propto (\d_\mu\phi)^2$, but 
$u^\mu$ is normalized, whereas $\rho $ is given by the norm of $(\d_\mu\phi)$, so they are independent quantities. Also note that
(\ref{etas}) implies $\eta/s\rightarrow 0$, so the resulting fluid is ideal.

\subsection{Ideal DBI hydrodynamics}

We now want to generalize the previous subsection to the case of the DBI model. As we saw, we cannot use the Belinfante tensor (\ref{belinDBI})
for hydrodynamics, since it does not become conformal in the free massless limit. We need to find an improved energy-momentum tensor 
that would reduce on-shell to the conformal tensor in the  free massless limit. 

For simplicity, we will first consider the $m\rightarrow 0$ limit, and we will then add the mass, considered as a perturbation. 
By comparison with the free massless scalar, we are led to choose the improved tensor
\bea
T_{\mu\nu}^I&=&T_{\mu\nu}^B+\d^\rho J_{\rho \mu\nu}\cr
J_{\rho\mu\nu}&=&+\frac{1}{6}\frac{(g_{\mu\nu}\d_\rho -g_{\mu\rho}\d_\nu)\phi^2}{\sqrt{1+\b^2(\d_\lambda\phi)^2}}.
\eea
Then we see that it can be rewritten as
\bea
T_{\mu\nu}^I&=&\frac{\frac{2}{3}(\d_\mu\phi\d_\nu\phi-g_{\mu\nu}(\d_\rho\phi)^2)-g_{\mu\nu}\b^{-2}}{\sqrt{1+\b^2(\d_\lambda\phi)^2}}
+\frac{g_{\mu\nu}}{\b^2}\cr
&&-\frac{\phi}{3\sqrt{1+\b^2(\d_\lambda\phi)^2}}
\left(\d_\mu\d_\nu\phi-\frac{\b^2\d_\nu\phi (\d_\mu\d^\rho\phi)\d_\rho\phi}{1+\b^2(\d_\lambda\phi)^2}\right)\cr
&&+g_{\mu\nu}\frac{\phi}{3}\d_\rho\frac{\d_\rho\phi}{\sqrt{1+\b^2(\d_\lambda\phi)^2}}\;,
\eea
and by using the equation of motion
\be
\d_\rho\frac{\d_\rho\phi}{\sqrt{1+\b^2(\d_\lambda\phi)^2}}=0\;,
\ee
we obtain the result,
\bea
T_{\mu\nu}^I&=&\frac{\frac{2}{3}(\d_\mu\phi\d_\nu\phi-g_{\mu\nu}(\d_\rho\phi)^2)-g_{\mu\nu}\b^{-2}}{\sqrt{1+\b^2(\d_\lambda\phi)^2}}
+\frac{g_{\mu\nu}}{\b^2}\cr
&&-\frac{\phi}{3\sqrt{1+\b^2(\d_\lambda\phi)^2}}
\left(\d_\mu\d_\nu\phi-\frac{\b^2\d_\nu\phi (\d_\mu\d^\rho\phi)\d_\rho\phi}{1+\b^2(\d_\lambda\phi)^2}\right).\label{improvDBI}
\eea
This result is easily seen to be traceless in the $\b\rightarrow 0$ limit.

We now make the same identification of the 4-velocity $u^\mu$, and for the same reasons as in the free case,
\be
u^\mu=\frac{\d_\mu\phi}{\sqrt{-(\d_\rho\phi)^2}}.
\ee
Now additionaly, we note that near $s=0$, as we have seen $\b^2(\d_\rho\phi)^2\rightarrow -1$, so 
\be
u^\mu\rightarrow \b^2\d_\mu\phi.
\ee

We can then identify in the improved tensor (\ref{improvDBI}) the first line as the ideal fluid tensor, and the second as a shear viscosity contribution. 
Then we obtain
\bea
P+\rho&=&-\frac{2(\d_\lambda\phi)^2}{3\sqrt{1+\b^2(\d_\mu\phi)^2}}\cr
P&=&\frac{1}{\b^2}(1-\sqrt{1+\b^2(\d_\lambda\phi)^2})+\frac{(\d_\rho\phi)^2}{3\sqrt{1+\b^2(\d_\lambda\phi)^2}}\;,
\eea
so the energy density and pressure are
\bea
\rho&=&\frac{1}{\b^2\sqrt{1+\b^2(\d_\lambda\phi)^2}}-\frac{1}{\b^2}\cr
P&=&\frac{1}{\b^2}(1-\sqrt{1+\b^2(\d_\lambda\phi)^2})+\frac{(\d_\rho\phi)^2}{3\sqrt{1+\b^2(\d_\lambda\phi)^2}}\;,
\eea
which leads at the shock, where $\b^2(\d_\mu\phi)^2\rightarrow -1$, to 
\be
\frac{3P+\rho}{\rho}\rightarrow 0. 
\ee
But that is the opposite of tracelessness, $P\simeq -\rho/3$. 

However, if we abandon the requirement of tracelessness in the $\b\rightarrow 0$ limit, we can choose instead the energy-momentum tensor 
with the opposite sign for $J_{\mu\nu\lambda}$,
\be
J_{\rho\mu\nu}=-\frac{1}{6}\frac{(g_{\mu\nu}\d_\rho -g_{\mu\rho}\d_\nu)\phi^2}{\sqrt{1+\b^2(\d_\lambda\phi)^2}}\;,
\ee
which leads after the use of the equations of motion to the improved tensor
\bea
T_{\mu\nu}^I&=&\frac{\frac{4}{3}(\d_\mu\phi\d_\nu\phi-g_{\mu\nu}(\d_\rho\phi)^2)-g_{\mu\nu}\b^{-2}}{\sqrt{1+\b^2(\d_\lambda\phi)^2}}
+\frac{g_{\mu\nu}}{\b^2}\cr
&&+\frac{\phi}{3\sqrt{1+\b^2(\d_\lambda\phi)^2}}
\left(\d_\mu\d_\nu\phi-\frac{\b^2\d_\nu\phi (\d_\mu\d^\rho\phi)\d_\rho\phi}{1+\b^2(\d_\lambda\phi)^2}\right).\label{improvDBIf}
\eea
From it, we extract the energy and pressure
\bea
\rho&=&\frac{1}{\b^2\sqrt{1+\b^2(\d_\lambda\phi)^2}}-\frac{1}{\b^2}\cr
P&=&\frac{1}{\b^2}(1-\sqrt{1+\b^2(\d_\lambda\phi)^2})-\frac{(\d_\rho\phi)^2}{3\sqrt{1+\b^2(\d_\lambda\phi)^2}}\;,
\eea
which now near the shock, when $\b^2(\d_\mu\phi)^2\rightarrow -1$, obey the conformal relation
\be
3P\simeq \rho\simeq \frac{1}{\b^2\sqrt{1+\b^2(\d_\mu\phi)^2}}\rightarrow \infty.
\ee

The justification for this choice of improved tensor is that the system at $s\rightarrow 0$, the shock, corresponds to an ultrarelativistic particle source, 
moving on $s=0$, and its energy-momentum tensor should be therefore conformal (it has energies $\gg \b^{-1/2}$).

Then the last term in the improved energy-momentum tensor is identified, as in the last subsection, with a shear viscosity contribution. Isolating the symmetric 
part of $\d_\mu u_\nu$, we find the shear viscosity
\be
\eta=\frac{\phi\sqrt{-(\d_\lambda\phi)^2}}{3\sqrt{1+\b^2(\d_\mu\phi)^2}}\;,
\ee
and considering that near the shock $\phi\propto \sqrt{s}\rightarrow 0$, we have
\be
\frac{\eta}{\rho+P}=\frac{\phi}{4\sqrt{-(\d_\lambda\phi)^2}}\rightarrow 0\;,
\ee
therefore the DBI model really describes an ideal hydrodynamics. 

The extension of the analysis here to the case of a mass term in the DBI action can be done straightforwardly. 

Introducing the mass, and considering the improved energy-momentum tensor
\bea
T_{\mu\nu}^I&=&T_{\mu\nu}^B+\d^\rho J_{\rho \mu\nu}\cr
J_{\rho\mu\nu}&=&+\frac{1}{6}\frac{(g_{\mu\nu}\d_\rho -g_{\mu\rho}\d_\nu)\phi^2}{\sqrt{1+\b^2[(\d_\lambda\phi)^2+m^2\phi^2]}}\;,
\eea
it can be rewritten as
\bea
T_{\mu\nu}^I&=&\frac{\frac{2}{3}(\d_\mu\phi\d_\nu\phi-g_{\mu\nu}(\d_\rho\phi)^2)-g_{\mu\nu}\b^{-2}(1+\b^2m^2\phi^2)}
{\sqrt{1+\b^2[(\d_\lambda\phi)^2+m^2\phi^2]}}
+\frac{g_{\mu\nu}}{\b^2}\cr
&&-\frac{\phi}{3\sqrt{1+\b^2[(\d_\lambda\phi)^2+m^2\phi^2]}}
\left(\d_\mu\d_\nu\phi-\b^2\d_\nu\phi\frac{(\d_\mu\d^\rho\phi)\d_\rho\phi+m^2\phi\d_\mu\phi}{1+\b^2[(\d_\lambda\phi)^2+m^2\phi^2]}\right)\cr
&&+g_{\mu\nu}\frac{\phi}{3}\d_\rho\frac{\d_\rho\phi}{\sqrt{1+\b^2[(\d_\lambda\phi)^2+m^2\phi^2]}}\;,
\eea
By using the equation of motion
\be
\d_\rho\frac{\d_\rho\phi}{\sqrt{1+\b^2(\d_\lambda\phi)^2+m^2\phi^2}}-m^2\frac{\phi}{\sqrt{1+\b^2(\d_\lambda\phi)^2+m^2\phi^2}}=0\;,
\ee
we obtain the result
\bea
T_{\mu\nu}^I&=&\frac{\frac{2}{3}(\d_\mu\phi\d_\nu\phi-g_{\mu\nu}(\d_\rho\phi)^2)-g_{\mu\nu}\b^{-2}\left(1-\frac{2}{3}\b^2m^2\phi^2\right)}
{\sqrt{1+\b^2(\d_\lambda\phi)^2+m^2\phi^2}}
+\frac{g_{\mu\nu}}{\b^2}\cr
&&-\frac{\phi}{3\sqrt{1+\b^2(\d_\lambda\phi)^2+m^2\phi^2}}
\left(\d_\mu\d_\nu\phi-\d_\nu\phi\frac{\b^2(\d_\mu\d^\rho\phi)\d_\rho\phi+m^2\phi\d_\mu\phi}{1+\b^2(\d_\lambda\phi)^2+m^2\phi^2}\right).\cr
&&
\eea

The same arguments apply as in the last subsection. The fluid is described in terms of $\phi$ only, but $u_\mu$ is defined in terms of the 
normalized $\d_\mu\phi$, and $\rho $ and $P$ in terms of its norm. In order to introduce a nonzero viscosity at the shock, we need to introduce second
derivatives of $\phi$ in the action.

\subsection{Viscous DBI hydrodynamics}

We can finally introduce a finite viscosity in the model, by adding a term inside the square root in the DBI action. 
We note that the term
\be
\sqrt{-(\d_\lambda\phi)^2}\d^\mu u_\mu=\sqrt{-(\d_\lambda\phi)^2}\d^\mu\left[\frac{\d_\mu\phi}{\sqrt{-(\d_\nu\phi)^2}}\right]
=\d^2\phi-\frac{(\d_\mu\phi)(\d^\mu\d^\rho\phi)(\d_\rho\phi)}{(\d_\lambda\phi)^2}\label{term}
\ee
when inside a Lagrangean generates a term in the energy-momentum tensor proportional to its variation under $g^{\mu\nu}$, i.e. to
\be
\d_\mu\d_\nu \phi-2\frac{(\d_{(\mu}\phi)(\d_{\nu)}\d_\rho\phi)\d^\rho\phi}{(\d_\lambda\phi)^2}+(\d_{\mu}\phi)(\d_{\nu}\phi)\frac{(\d_\rho\phi)(\d^\rho\d^\sigma\phi)
(\d_\sigma\phi)}{[(\d_\lambda\phi)^2]^2}.
\ee
On the other hand, 
\bea
\sqrt{-(\d_\lambda\phi)^2}\frac{\d_{\mu}u_{\nu}+\d_\nu u_\mu}{2}&=&\d_\mu\d_\nu\phi-\frac{\d_{(\nu}\phi(\d_{\mu)}\d_\rho\phi)\d^\rho\phi}{(\d_\lambda\phi)^2}\cr
\sqrt{-(\d_\lambda\phi)^2}\frac{a_\mu u_\nu+a_\nu u_\mu}{2}&=&-\frac{\d_{(\nu}\phi(\d_{\mu)}\d_\rho\phi)\d^\rho\phi}{(\d_\lambda\phi)^2}
+(\d_{\mu}\phi)(\d_{\nu}\phi)\frac{(\d_\rho\phi)(\d^\rho\d^\sigma\phi)(\d_\sigma\phi)}{[(\d_\lambda\phi)^2]^2}\;,\cr
&&
\eea
so the term in the energy-momentum tensor coming from the variation of (\ref{term}) is proportional to 
\be
\sqrt{-(\d_\lambda\phi)^2}\left[\frac{\d_{\mu}u_{\nu}+\d_\nu u_\mu}{2}+\frac{a_\mu u_\nu+a_\nu u_\mu}{2}\right]\;,
\ee
leading to the correct $\pi_{\mu\nu}^{(1)}$, with a nonzero $\eta$, and $\zeta-2\eta/3=0$.

On the $\phi=\phi(s)$ solution, we find
\bea
\d^2\phi&=&-4\frac{d}{ds}\left(s\frac{d\phi}{ds}\right)\cr
(\d_\mu\phi)(\d^\mu\d^\nu\phi)\d_\nu\phi&=&8s\phi'^2(2s\phi''+\phi')\cr
\d^2\phi-\frac{(\d_\mu\phi)(\d^\mu\d^\nu\phi)\d_\nu\phi}{(\d_\lambda\phi)^2}&=&-2\phi'\;,
\eea
and moreover, on the solution near $s=0$, $\phi(s)\simeq \b^{-1}\sqrt{s}$, we have $\b^2(\d_\lambda\phi)^2\rightarrow -1$.

We can consider therefore the modified DBI action 
\be
S=-\b^{-2}\int d^4x \left[\sqrt{1+\b^2[(\d\phi)^2+m^2\phi^2]+\a\left[\d^2\phi-\frac{(\d_\mu\phi)(\d^\mu\d^\rho\phi)(\d_\rho\phi)}{(\d_\lambda\phi)^2}\right]}
-1\right]\;,\label{DBImod}
\ee
which on the $\phi=\phi(s)$ solution becomes
\be
S=-\b^{-2}\int d^4x \left[\sqrt{1+\b^2\left(-4s\left(\frac{d\phi}{ds}\right)^2+m^2\phi^2\right)+\a\left(-2\frac{d\phi}{ds}\right)}-1\right]\;,\label{DBImods}
\ee
and it will give a shear viscosity of 
\be
2\eta=\frac{\a \sqrt{-(\d_\lambda\phi)^2}}{\b^2\sqrt{1+\b^2[(\d\phi)^2+m^2\phi^2]+\a\left[\d^2\phi-\frac{(\d_\mu\phi)(\d^\mu\d^\rho\phi)
(\d_\rho\phi)}{(\d_\lambda\phi)^2}\right]}}\rightarrow \infty\;,
\ee
which blows up near the shock, together with a nonzero bulk viscosity $\zeta=2\eta/3$. 

The resulting fluid will, of course, {\em satisfy the relativistic Navier-Stokes equation (\ref{relNavSto}).}

But it doesn't matter that the shear viscosity is infinite, since only the ratio $\eta/s$, with $s$ the entropy density, matters, and $s$ itself is infinite
near the shock. From the thermodynamic relation $U+PV-TS=0$, we obtain that 
\be
\rho+P=Ts\Rightarrow \frac{\eta}{s}=\frac{\eta T}{\rho +P}\;,
\ee
and we have seen in the previous subsection that $\eta/(\rho+P)\rightarrow 0$ in the case of the pure DBI action. With the extra term, we obtain
\be
\frac{\eta}{s}=T\frac{\a}{\b^2\sqrt{-(\d_\lambda\phi)^2}}\rightarrow T\frac{\a}{\b}\;,
\ee
i.e. a finite result. Note that here $T$ is the temperature of the (fluid) fireball. Moreover, we see that we obtain the relation
\be
\frac{1}{T}\frac{\eta}{s}=\frac{\a}{\b^2}\label{thermod}
\ee
that relates the thermodynamic quantities to the parameters in the action. 

It remains to fix the parameter $\a$ from experiments. It would be ideal to determine it from the hydrodynamics description, but as in the 
case of the parameter $\b$, it is not clear how to do that. Instead, one must look again at the action as an effective action in a chiral perturbation 
theory, and think of using usual scattering experiments.

Since for small $\phi$ it generates a term in the Lagrangean expanded in perturbation theory of 
\be
\sim \frac{\a^2}{\b^2}(\d^2\phi)^2\sim \frac{\a^2 m_\pi^4}{\b^2}\phi^2\;,
\ee
where we have used that $\d^2\phi\sim m_\pi^2\phi$ for small $\phi$ (this is the free equation of motion), we expect 
\be
\frac{\a^2 m_\pi^2}{\b^2}\sim 1\Rightarrow \a\sim \frac{\b}{m_\pi}=\frac{\sqrt{2}}{\sqrt{3}f_\pi m_\pi^2}.
\ee
Note that this relation is only in order of magnitude, since we already have a mass term in the action, but this term would be of the same order.
Moreover, from this together with (\ref{thermod}), we obtain the condition 
\be
T\sim m_\pi \frac{\eta}{s}\propto m_\pi\;,
\ee
since experimentally $\eta/s$ is close to the universal value of $1/(4\pi)$ valid for most models which have a AdS/CFT gravity dual with a 
black hole, hence can be considered constant. It is interesting that the same relation $T\propto m_\pi$ was 
obtained in \cite{Nastase:2005rp}. Note however that numerically, in order to match the experimentally measured value of the temperature $T$, 
one must have something very close to the relation conjectured in 
\cite{Nastase:2005rp}, namely $T\simeq 4 m_\pi/\pi$, leading from (\ref{thermod}) to 
\be
\a\simeq\b/(16 m_\pi)\label{instead}
\ee
instead.

\subsection{Viscosity as regulator of the shock}

The equation of motion of the action (\ref{DBImods}) is (after making some relevant cancellations)
\be
4\frac{d}{ds}\left(s\frac{d\phi}{ds}\right)+m^2\phi=\frac{8\b^2s \phi'^2(\phi'+m^2\phi)+3\a\phi'(2\phi'+m^2\phi)-\a^2\b^{-2}\phi''}{1+\b^2m^2\phi}.\label{visceq}
\ee
We can check that now the solution $\phi(s)\simeq \b^{-1}\sqrt{s}$ is not valid anymore as $s\rightarrow 0$, which is not surprising, since the 
new term added to the action will dominate at the highest energies, or equivalently at the shortest distances; 
neither is a general power law $As^\a$ a solution. The solution 
\be
\phi\simeq \frac{\a(\sqrt{5}-3)}{4\b^2}\ln s
\ee
would work, but cannot be made to be continuous with $\phi=$constant at $s<0$. Instead, a Taylor expanded solution
\be
\phi(s)\simeq A+Bs+Cs^2+...
\ee
can, and leads to the condition
\be
4B+m^2A=\frac{3\a B(2B+m^2A)-2\a^2\b^{-2}C}{1+\b^2m^2A^2}
\ee
that fixes $C$ in terms of $A$ and $B$. This solution is valid as long as 
\be
\sqrt{s}\ll\frac{3\a}{4\b}\sim\frac{1}{m_\pi}\;,
\ee
which is the condition that the term with $\a$ in the equation of motion (\ref{visceq}) dominates over the left-hand side for the $\sqrt{s}/\b$ 
solution,
so it would seem to replace the solution $\phi\simeq \b^{-1}\sqrt{s}$ valid at $\a=0$.  But if we have (\ref{instead}), we 
would in fact have a window of $\sqrt{s}$,
\be
\frac{3\a}{4\b}\ll \sqrt{s}\ll \frac{1}{m_\pi}
\ee
where the solution $\phi\simeq \sqrt{s}/\b$ would still be valid. 

Note then that the introduction of the extra term in the square root in the action has the effect of modifying the solutions at the highest energies, or the 
smallest distances, which gives a regularization of the behaviour of the energy density, since now 
\be
\b^2(\d_\mu\phi)^2=-\b^24s\phi'^2\simeq -4\b^2s B^2\rightarrow 0\;,
\ee
which is still negative, but very small, so the $1/\sqrt{1+...}$ factor in the energy density is now finite, as opposed to the case with $\a=0$, when 
it became infinite at $s=0$.

\section{Conclusions}

In this paper I have presented a way to describe the QGP hydrodynamics as arising from a modification of the Heisenberg model for 
high energy nucleon-nucleon scattering, with a massive DBI action for a scalar pion. The behaviour of the quantum field theory in the 
classical regime of high multiplicity, giving a classical field, can be described as a derivative expansion for a fluid velocity variable, 
i.e. a hydrodynamics expansion. The fluid is well defined near a shock in the (1+1)-dimensional shockwave classical solution. It would 
be interesting to extend the analysis to a more realistic (3+1)-dimensional shockwave, but we leave that for future work.

The DBI model gives a relativistic, ideal and approximately conformal fluid near the shock, but one can introduce shear viscosity, 
as well as bulk viscosity, by 
adding an extra term inside the square root of the DBI action. I have found that the presence of this term regulates the behaviour of the 
energy density near the shock, making it finite, as required physically. The resulting viscous fluid satisfies the relativistic Navier-Stokes equation.
The velocity $u^\mu$ was given by the normalized $\d^\mu\phi$, $\rho$ and $P$ in terms of the norm of $\d^\mu\phi$, and to have
viscosity we needed the $\d^2\phi$ term in the action.

The parameters of the model, $\a$ and $\b$, were not fixed from the hydrodynamics description, but rather by assuming that the action is
also valid in low multiplicity scatterings, as a kind of chiral perturbation theory. That fixed $\b$, and $\a$ was fixed only as an order of 
magnitude, since it generates a term that is degenerate with a mass term already in the Lagrangean. 
I have then found the thermodynamics relation $\eta/(sT)=\a/\b^2$, which led to $T\propto m_\pi$. 
It would be very interesting to see if one can fix more precisely $\a$ and $\b$, and thus obtain a precise prediction for $\eta/(sT)$. 
One should also explore whether there is a way to fix them independently from the hydrodynamics, though as I mentioned, I could find no way.

{\bf Acknowledgements}

I would like to thank Stefano Finazzo for useful comments and references and to Jacob Sonnenschein for suggestions on the manuscript. 
My work is supported in part by CNPq grant 301709/2013-0 and FAPESP grant  2014/18634-9.

\bibliographystyle{utphys}
\bibliography{heisenberg2}

\end{document}